\documentclass[aps,prd,twocolumn,superscriptaddress,nofootinbib,showpacs]{revtex4-1}

\usepackage[pdftex]{graphicx}
\usepackage{epsfig}
\usepackage{amsmath}
\usepackage{amssymb}
\usepackage{dcolumn}% Align table columns on decimal point
\usepackage{multirow}

\newcommand{\beq}{\begin{equation}}
\newcommand{\eeq}{\end{equation}}
\newcommand{\bear}{\begin{eqnarray}}
\newcommand{\eear}{\end{eqnarray}} \newcommand{\ba}{\begin{array}}
\newcommand{\ea}{\end{array}}
\newcommand{\lae}{\begin{array}{c}\,\sim\vspace{-1.7em}\\< 
\end{array}}
\newcommand{\gae}{\begin{array}{c}\,\sim\vspace{-1.7em}\\> 
\end{array}}

\begin{document}

%\linenumbers
\title{Indirect Probes of Supersymmetry Breaking in Multi-Km$^3$ Neutrino Telescopes} 
\author{Ivone Freire M. Albuquerque} 
\email{ifreire@if.usp.br}
\author{Jairo Cavalcante de Souza} 
\email{jairocavalcante@gmail.com}
\affiliation{Instituto de F\'{i}sica, Universidade de S\~ao Paulo, S\~ao Paulo, Brazil}

\date{\today}

\begin{abstract}
Recently it has been shown \cite{aceuso} that fluorescence telescopes with a large
field of view can indirectly probe the scale of supersymmetry
breaking. Here we show that depending on their ability to fight a large background, 
multi-Km$^3$ volume neutrino telescopes might independently probe a similar breaking
scale region, which lies between $\sim 10^5$ and $\sim 5
\times 10^6$~GeV. The scenarios we consider have the gravitino as the lightest supersymmetric particle,
and the next to lightest (NLSP) is a long lived
slepton.  Indirect probes complement a proposal \cite{abc} that demonstrates 
that 1~Km$^3$ telescopes can directly probe this breaking scale. A
high energy flux of neutrinos might interact in the Earth producing
NLSPs which decay into taus. We estimate the rate of taus, taking
into account the regeneration process, and the rate of secondary
muons, which are produced in tau decays, in   
multi-km$^3$ detectors. 
\end{abstract}


\maketitle

\section{\label{intro} Introduction}

Extensions of the standard model of particle physics (SM) are currently
being probed at the Large Hadron Collider (LHC). At the same time,
features of these extensions can be probed by non accelerator
experiments. It has recently been shown~\cite{aceuso}
that the supersymmetry breaking scale ($\sqrt{F}$) can be probed
indirectly by large fluorescence telescopes, such as the JEM-EUSO
Observatory. This scale, as well as Universal Extra Dimensions
scenarios~\cite{abkk}, can also be directly probed by 1~Km$^3$ volume neutrino telescopes~\cite{abc,abcd}, such
as IceCube. Here we show that depending on their ability to fight a large background, 
multi-Km$^3$ volume neutrino telescopes might also indirectly probe
the scale of supersymmetry breaking.

We investigate scenarios where supersymmetry (susy) is broken at
$\sqrt{F}<10^{10}$~GeV , and the  lightest supersymmetric particle
is the gravitino. In many of these scenarios, including Gauge
Mediation Susy Breaking models~\cite{gmsb}, the NLSP is a right-handed
stau, whose lifetime is~\cite{abc}:

\beq
 c\tau = \left(\frac{\sqrt{F}}{10^7{\rm~GeV}}\right)^4\, 
\left(\frac{100~{\rm GeV}}{m_{\tilde{\tau}_R}}
\right)^5\,10~{\rm km}~,
\label{eq:ctau}
\eeq
where ${m_{\tilde\tau_R}}$ is the stau mass. The NLSP range depends on
$\sqrt{F}$, and for $\sqrt{F} \gae 5 \times 10^6$, they  travel a long way before decaying while for lower
values they decay after a short travel. It has
been shown~\cite{abc,abcd} that neutrino telescopes can directly probe
this scale, probing scenarios where $5 \times 10^6 \lae \sqrt{F} \lae
5 \times 10^{8}$~GeV, while large fluorescence telescopes~\cite{aceuso,tese}, such as the JEM-EUSO
Observatory~\cite{euso}, can probe scenarios where NLSPs decay and
$\sqrt{F} \lae 5 \times 10^6$~GeV.

In this article we show that multi-Km$^3$ volume neutrino telescopes can probe
the same $\sqrt{F}$ region as large fluorescence telescopes. NLSP
decays will yield a few events per year in the projected KM3Net 
neutrino telescope ~\cite{kmnet}, or in an expanded version of IceCube~\cite{icplus}.
However, while fluorescence telescopes have
a clean signature of these events, probes by neutrino telescopes will depend on the ability
to discriminate taus ($\tau$s)  produced in NLSP decays from a large background.

A detectable rate of NLSPs in a 1 Km$^3$ neutrino telescope, such a
IceCube, can be produced by high energy neutrinos interactions in the
Earth. Here we consider the same NLSP production
mechanism, but analyze scenarios where these decay, probing 
a complementary breaking scale, where $\sqrt{F} \lae 5 \times 10^{6}$~GeV.

High energy $\tau$s are produced in NLSP decays and undergo a
regeneration process, where $\tau$ neutrinos ($\nu_\tau$) produced from $\tau$
decay, charge current interact in the Earth and produce a new $\tau$. This process will expand the 
fraction of the Earth volume within which the $\tau$ can decay and still reach the detector.
We also consider the flux of muons produced when these secondary
$\tau$s decay. 

The caveat in probing $\sqrt{F}$ \cite{abc} comes from the fact that
NLSPs lose  less energy and have a much larger
range while transversing the Earth when compared to
SM leptons. This larger range compensates its lower production cross
section. Here, when NLSP decays are
considered, an extra decaying volume is gained due to the $\tau$ regeneration process. 

Multi-km$^3$ neutrino telescopes proposals exist, either as a new
telescope in the Mediterranean Sea \cite{kmnet} or  as an expansion of
IceCube~\cite{icplus} . Here we show that depending on their final
design and their ability to fight a large background, they might be able to probe the $\sqrt{F}$
region which allows NLSPs decays in the Earth.

In the next section we describe our simulation of
NLSP production, propagation and energy loss, which reproduces the
analysis described in~\cite{abc,abcd}. Follows the description of
our simulation of NLSP decays with $\tau$ production; $\tau$ decay and regeneration process 
as well as muon production, which is the same used
in~\cite{aceuso}. The event rate in neutrino telescopes will depend on
its design. In Section~\ref{sec:rate} we discuss this point, and determine the rates of these events in
a 1 Km$^3$ neutrino telescope. In section~\ref{sec:multi} the rate
is determined for multi-km$^{3}$ telescopes. 
In Section~\ref{sec:bckg} we analyze $\tau$ specific signatures, as well as a cosmological 
background. Finally we present our conclusions.

\section{\label{sec:nlsp} NLSP PRODUCTION}

As shown in \cite{abc}, NLSPs can be produced from the interaction of
high energy neutrinos in the Earth. Although the production cross section related to this
process is even lower than the one for SM leptons, the flux of NLSPs
will still be considerable due to its much larger range, as described
in the previous section.

Here we take the Waxman-Bahcall (WB)~\cite{wb} upper limit on the
neutrino flux hitting the Earth as a reference for our analysis. Our
result is normalized by this upper limit, and can be translated
to any other flux if properly rescaled. We consider neutrino
production from optically thin sources, and the WB limit (for a
maximized cosmological evolution of sources) is given by
\beq
\left(\frac{d\phi_\nu}{dE_\nu}\right)_{\rm WB} = \frac{4 \times
10^{-8}}{E_\nu^2} {\rm GeV~ cm^{-2} s^{-1} ~sr^{-1}}~,
\label{wblimit}
\eeq
where $E_\nu$ is the neutrino energy.
It is important to note that although this limit was set for muon and
electron neutrinos, our analysis is independent of the initial
neutrino flavor. Any flavor of the left-handed sleptons, produced from neutrino interactions, immediately decay
producing a chain that will always end up with right-handed staus (the NLSPs).
For this reason, neither the initial neutrino flavor nor
cosmogenic mixing due to oscillations alter our results.

NLSPs will be produced by the interaction of cosmogenic neutrinos in
the Earth and the production cross section is determined as
in~\cite{abcd}. It follows the same pattern as for SM
lepton production by charge current interactions, where now the
up or down type quark is a squark ($\tilde{q}$) and the chargino is the
mediator in the t-channel. We also account for the sub-dominant
neutralino exchange. From this interaction a left-handed slepton ($\tilde{l}_L$) and a
$\tilde{q}$ are produced and promptly decay, always producing two NLSPs 
at the end of the decay chain. The production energy threshold
depends on $m_{\tilde{q}}$ and $m_{\tilde{l}_L}$ and is $\sim
10^5$~GeV, and the NLSP initial energy
 is $\sim E_{\nu}/6$~\cite{abcd} . These NLSPs are typically
right handed staus ($\tilde{\tau}_R$) which travel in parallel through
the Earth. We take the chargino mass as $m_{\tilde
  w}=250$~GeV,  and $m_{\tilde{l}_L}=150$~GeV,
$m_{\tilde{\tau}_R}=250$~GeV for the left-handed stau and the NLSP
respectively, and consider three possibilities for the squark mass, $m_{\tilde q}=300$, ~$600$ or $900$~GeV.
There are constrains to the $\tilde{\tau}_R$ from big-bang nucleosynthesis~\cite{bbn}. 

NLSP production depends on the probability that neutrinos will
interact  in the Earth, and we use the Earth density profile model described in
\cite{gandhi,earth}. Our simulation of
NLSP propagation includes its energy loss, due both to ionization and
radiative processes. It follows the analysis described in~\cite{abcd,ina},
where it is shown that the NLSP energy degradation  due to radiative losses
is suppressed when compared to SM leptons. 

We check our NLSP production and propagation simulation by reproducing
the direct detection rate and the energy distribution shown in \cite{abc,abcd}.

\section{\label{sec:rate} NLSP Decay and Event Rates in 1 Km$\mathbf{^3}$ Neutrino Telescopes}

\begin{figure}
%\vspace*{-1.5cm}
\includegraphics[width=\linewidth,height=7.5cm]{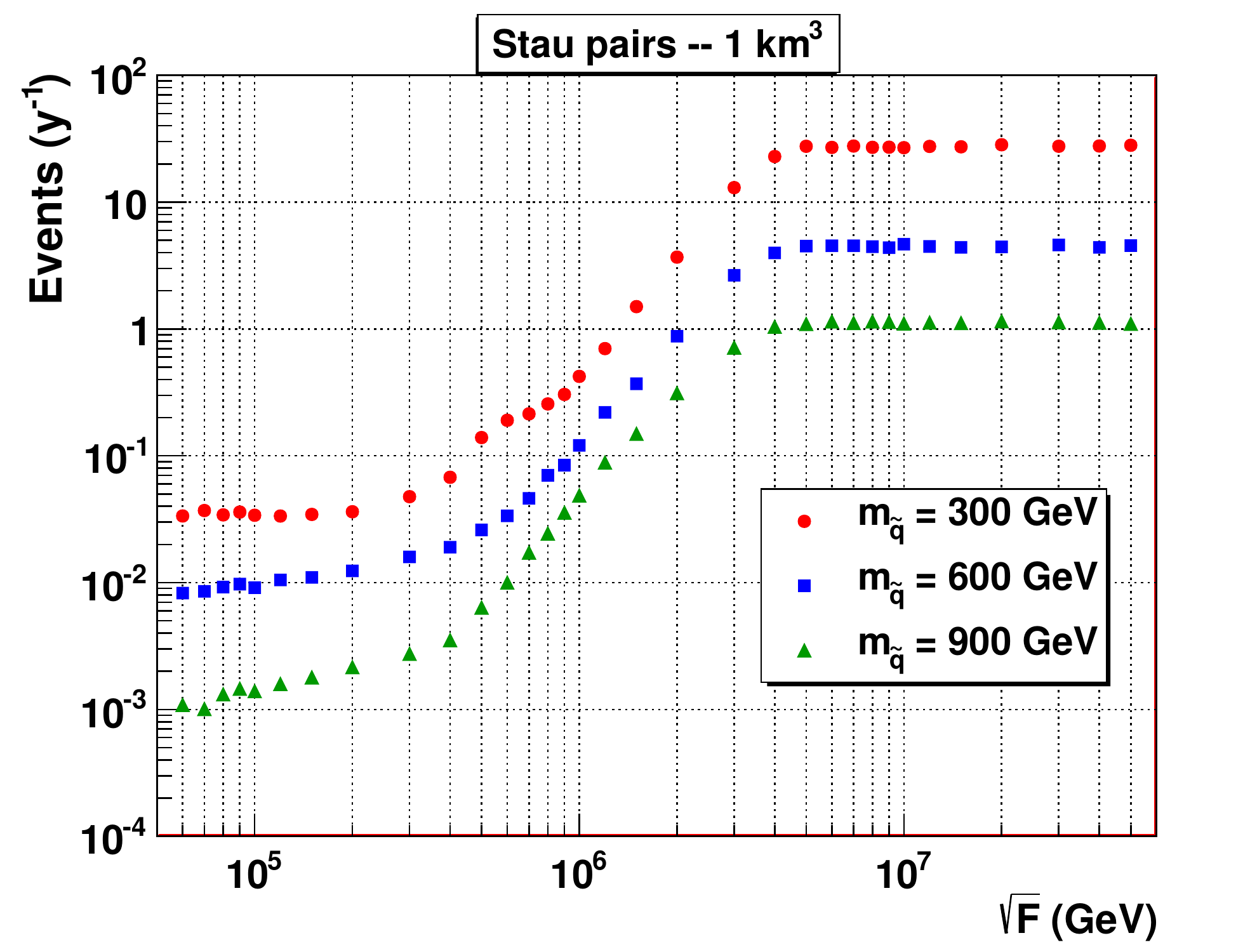}
%\vspace*{-1.cm}
\caption{Number of NLSP events per year in a km$^3$ neutrino telescope versus
the scale of susy breaking $\sqrt{F}$. It is assumed an 
initial neutrino flux equivalent to the WB limit. The curves are shown for
different $m_{\tilde q}$ values.}
\label{fig:sqf}
\end{figure}

Here we consider scenarios where $\sqrt{F}$ is such that 
NLSPs decay after a short travel through the Earth. This distance equals
$\gamma c \tau$, and can be determined from Equation~\ref{eq:ctau}. They decay into a
$\tau$ and a gravitino, where the $\tau$ is much heavier than the gravitino.
Figure~\ref{fig:sqf} shows the number of NLSPs per year in a km$^3$
neutrino telescope (positioned $\sim 2$~Km deep in the Earth) versus $\sqrt{F}$, and it is clear that NLSPs
decay for $\sqrt{F} \lae 5 \times 10^{6}$~GeV. 

In our Monte Carlo simulation $\vartheta(10^5)$ neutrinos  isotropically distributed reach the Earth. These are generated with
energies between $10^5$ to $10^{12}$~GeV, and are normalized by the WB
limit. NLSPs are then produced according to the neutrino interaction
probability distribution, which depends on the Earth density profile and
the NLSP production cross section, and propagated, taking into account
their energy loss, as described in the previous section. The neutrino
interaction probability is convoluted by its survival probability to
account for SM lepton production.

Once the NLSP production and propagation is simulated, NLSPs are
randomly selected to decay according to their decay probability distribution, $P_d = 1 - \exp(-m x /E_{\tilde
  \tau_L} c \tau )$. The decay generate $\tau$s isotropically
distributed in the center of mass frame, which are boosted to the
laboratory frame, mainly following the same direction as the
NLSPs. $\tau$s will then decay and generate $\nu_\tau$s and a
regeneration process might take place. In 18\% of the $\tau$ decay, a muon will be
produced and, if close enough to the detector, its propagation will also be simulated.

\begin{figure}
%\vspace*{-1.5cm}
\includegraphics[width=\linewidth,height=7.5cm]{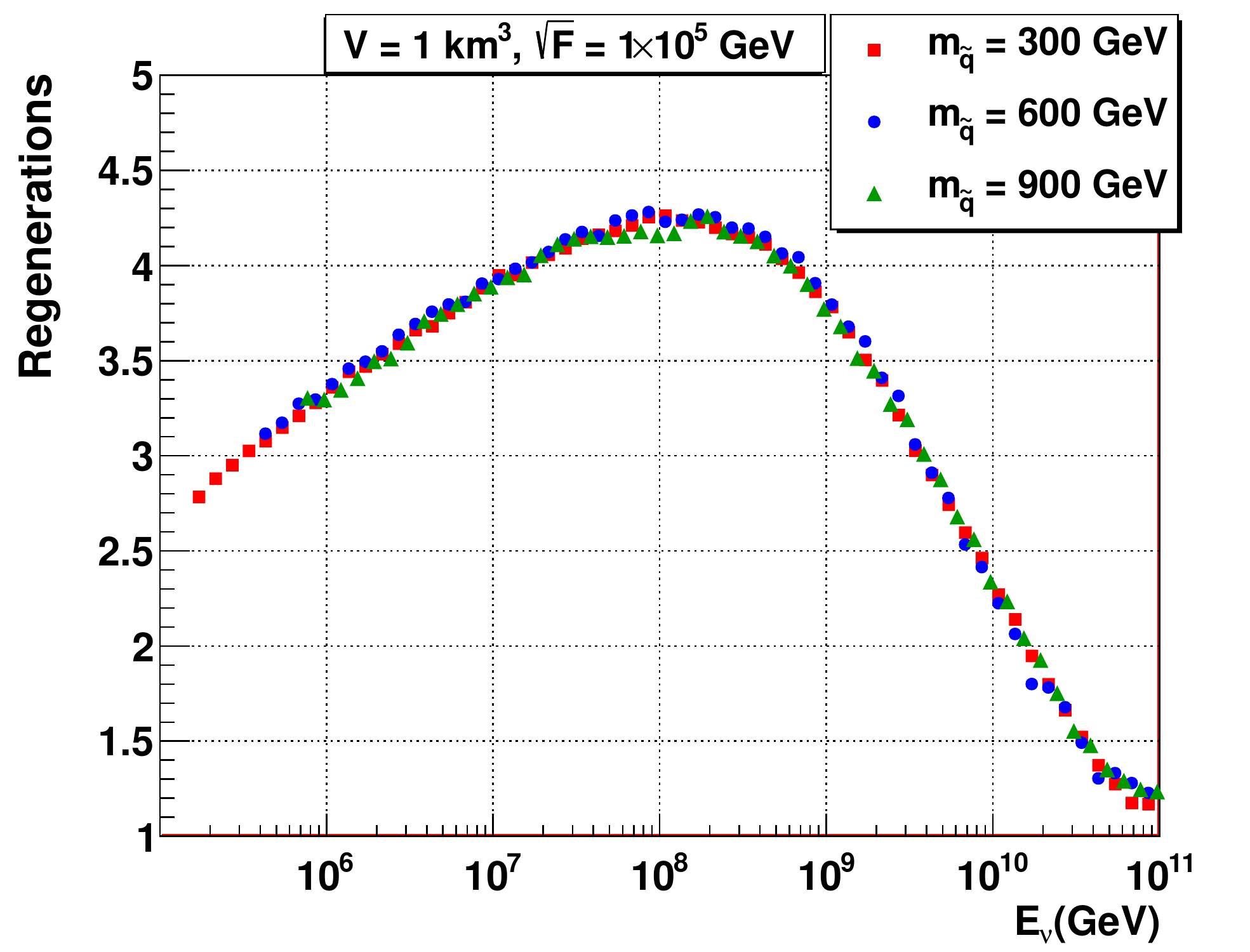}
\caption{Average number of $\tau$ regenerations. Only the $\tau$s which reach a 1 km$^3$ 
detector are taken into account. Results are shown for three values of $m_{\tilde q}$, and
as expected show no dependency on this parameter.}
\label{fig:reg}
\end{figure}

The regeneration process depends on the $\tau$ energy. The
fraction of energy carried by the $\nu_\tau$ is determined as
described in~\cite{crotty,dutta}. The average number of regenerations
are shown in Figure~\ref{fig:reg} as a function of the neutrino energy
$E_\nu$, for $\tau$s that reach a 1 Km$^3$ detector positioned as IceCube.

We note that two NLSPs are always 
produced from the neutrino interaction in the Earth. These NLSPs will decay at different
points in the Earth and the generated $\tau$s will not follow the same coincidence 
pattern in the detector as the NLSPs. 

In order to understand the possibility of NLSP indirect detection by
neutrino telescopes, the detector shape is
relevant. The $\tau$ average path and number of regenerations depend strongly on its
energy. Up to $10^6$~GeV the average path length is much smaller than a Km, 
but larger energies imply that this distance might be longer than the detector size.
Events arriving horizontally with respect to the detector can go through fewer regeneration 
processes than the ones coming from below. Therefore, it is important to determine the event 
rate as a function of the detector volume as well as to take into account the detector shape.
One important point is that due to the regeneration process, the effective volume for NLSP 
production will be enhanced when compared to direct NLSP detection.  

We consider a 1~km$^3$ cylindrical detector, with a 0.564~Km radius,
1~Km height, and buried 2~Km deep in the Earth. Our results are therefore relevant both for IceCube~\cite{icplus} and 
Km3Net~\cite{kmnet} neutrino telescopes. We show our results as a function of two arbitrary 
values of $\sqrt{F}$, $10^5$ and $10^7$~GeV, where the lower value represents the maximum 
NLSP decay rate and the upper one the region where NLSPs are stable 
while transversing the Earth.

\begin{figure}
%\vspace*{-1.5cm}
\includegraphics[width=\linewidth,height=7.5cm]{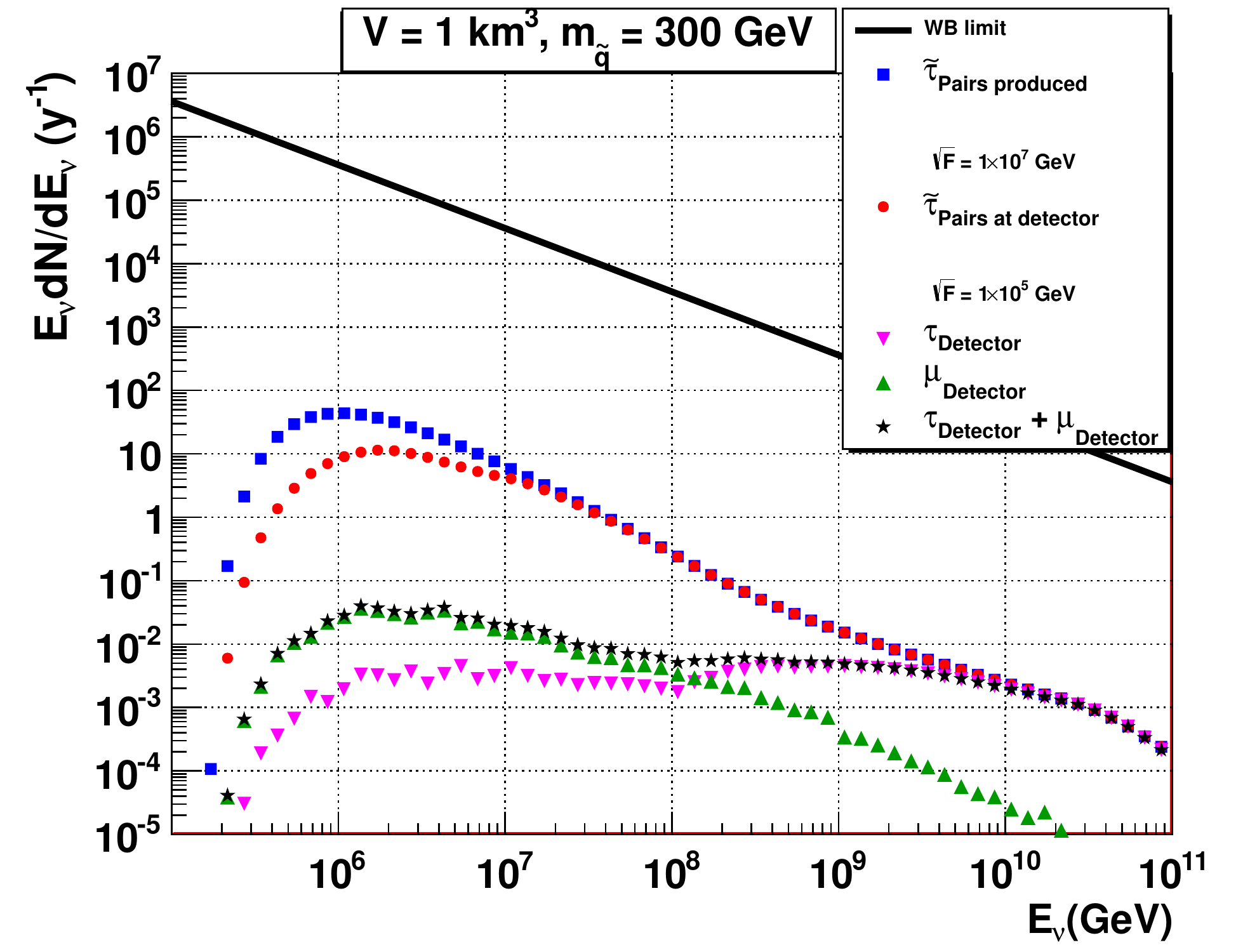}
\caption{Energy distribution of produced $\tilde{\tau}$ pairs (blue square), the fraction
of these that reach the detector (red dots) assuming $\sqrt{F} = 10^7$ GeV and of
$\tau$s (pink triangles) and $\mu$s (green triangles) as well as their sum (black stars) that reach 
the detector assuming $\sqrt{F} = 10^5$ GeV. Here we take $m_{\tilde q} = 300$~GeV. The
distributions for $m_{\tilde q} = 600$ and 900~GeV, are very similar in shape with lower
number of events.}
\label{fig:ev1km3}
\end{figure}

Figure~\ref{fig:ev1km3} shows the energy distribution for events reaching a 1 Km$^3$ detector.
Up to neutrino energies of $10^7$~GeV the 
$\tau$ rate is smaller than the $\mu$ one and this relation inverts at $\sim 10^8$~GeV.
At lower energies the $\tau$ decays almost instantly, which lower it's detection probability
when compared to larger energies where it travels for a while before decaying. For 
$E_\nu \geq 5 \times 10^9$~GeV all $\tau$s will reach the detector. This figure assumes
$m_{\tilde q} = 300$~GeV, whereas the  distributions for $m_{\tilde q} = 600$ and 900~GeV, 
are very similar in shape with lower number of events. The total rate of events
in a Km$^3$ detector per year, for the three values of $m_{\tilde q}$, are shown in 
Table~\ref{tab:ev1y}. It is important to note that the number 
of events in larger detectors will not necessarily scale linearly with these results, as we will 
show below, due to the importance of the shape of the detector.

\begin{table}
%vspace*{-2.5cm}
\begin{center}
\vspace{11pt}
\begin{tabular}{|c|c|c|c|c|c|}
\hline
$m_{\tilde q}$ (GeV) & V (km$^3$) & $\tilde{\tau}$ (y$^{-1}$) & $\tau$ (y$^{-1}$) & $\mu$ (y$^{-1}$) & $\mu+\tau$ (y$^{-1}$)   \\
\hline
300  &  1 &   27.7  &   0.07 & 0.21   & 0.28 \\
300  & 21 &  521.5  &   1.60 & 2.36   & 3.95 \\
300  & 65 & 1450.5  &   4.41 & 5.85   & 10.26 \\
\hline 
600  &  1 &    4.5  &   0.02 & 0.04   & 0.06 \\
600  & 21 &  100.6  &   0.51 & 0.47   & 0.98 \\
600  & 65 &  260.0  &   1.47 & 1.27   & 2.74 \\
\hline 
900  &  1 &    1.1  &   0.01 & 0.01   & 0.02 \\
900  & 21 &   28.2  &   0.22 & 0.15   & 0.37 \\
900  & 65 &   73.0  &   0.65 & 0.41   & 1.05 \\
\hline
\end{tabular}
\caption{$\tilde{\tau}$ pair, $\tau$ and $\mu$ rates per year in a Km$^3$ neutrino
telescope at $\sim 2$~Km deep, for neutrino telescopes of 1, 21 and 65 Km$^3$ volumes.
Results are given  for 3 $m_{\tilde q}$, assuming $\sqrt{F} = 10^7 (10^5)$~GeV
for $\tilde{\tau}$ pairs ($\tau$s and $\mu$s).}
\label{tab:ev1y}
\end{center}
\end{table}

\begin{figure}
%\vspace*{-1.5cm}
\includegraphics[width=\linewidth,height=7.5cm]{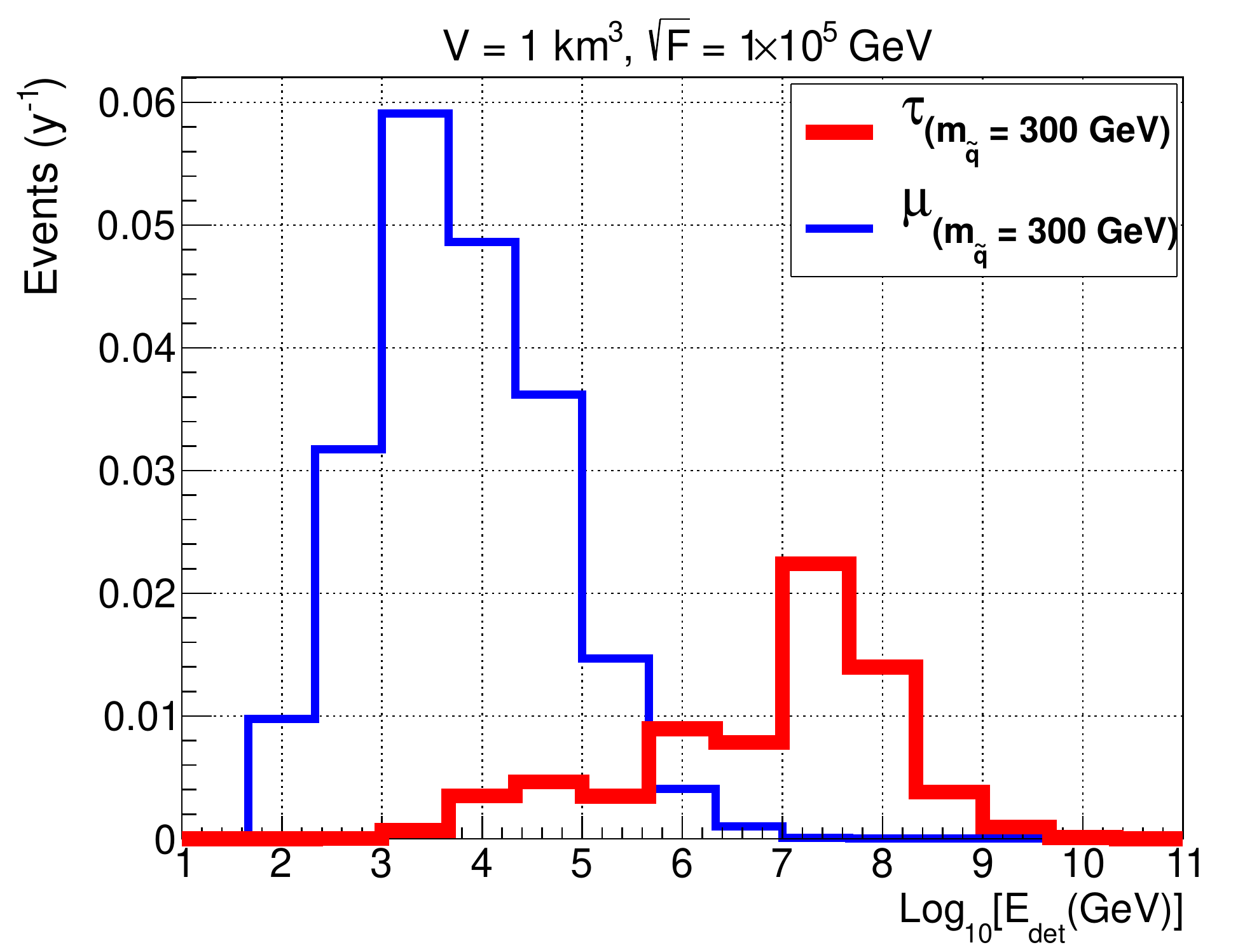}
\caption{Distribution of $\tau$ and $\mu$ energy at a 1~Km$^3$
  detector for $\sqrt{F} = 10^5$~GeV. Results are shown for $m_{\tilde
    q} = 300$~ GeV. Other values of $m_{\tilde q}$s have similar distributions
with less events.}
\label{fig:endet}
\end{figure}

The $\tau$ and $\mu$ energy at the detector is shown in Figure~\ref{fig:endet}. It can be
seen that most $\tau$s that reach the detector have larger energies
than the $\mu$s, reflecting the distribution in Figure~\ref{fig:ev1km3}.

\section{\label{sec:multi} DECAY RATES IN MULTI-Km$\mathbf{^3}$ TELESCOPES}

As can be seen in Table~\ref{tab:ev1y}, the indirect probe of the susy breaking scale
is not promising at 1 Km$^3$ neutrino telescopes. However there are studies~\cite{icplus}
and proposals~\cite{kmnet} of multi-Km$^3$ detectors. Here we determine the $\tau$ and $\mu$ rates from
NLSP decays in these detectors and analyze how well they can probe this scale.

We take a 2 and 4 Km radius extension to our current cylindrical configuration (as proposed in 
\cite{icplus} for an IceCube extension) as our case scenarios. This leads to 21 and 65 Km$^3$ 
volumes. It is important to note that due to the regeneration process, the increase in the 
indirect detection rate will depend on the shape of the extended detector. 

\begin{figure}
%\vspace*{-1.5cm}
\includegraphics[width=\linewidth,height=15.cm]{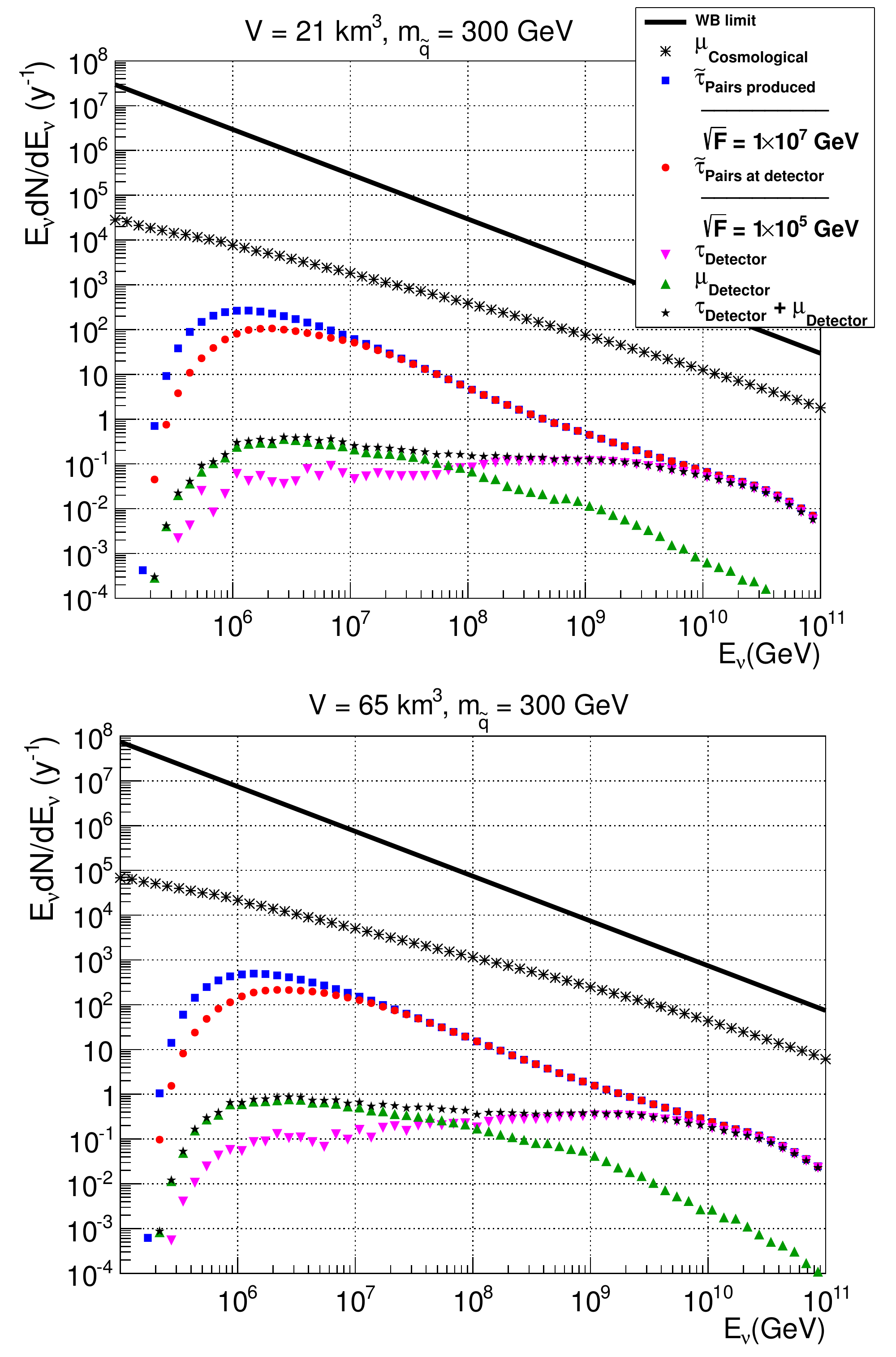}
%\includegraphics[width=\linewidth,height=9.5cm]{taus_21km3.pdf}
%\vspace*{-.5cm}[H]
%\includegraphics[width=\linewidth,height=18.5cm]{nucosm_65km3.pdf}
%\includegraphics[width=\linewidth,height=9.5cm]{taus_65km3.pdf}
\caption{Same as Figure~\ref{fig:ev1km3}, but now for extended neutrino telescopes. Top
plot shows results for 21~Km$^3$ and bottom for 65~Km$^3$ volume telescopes. Both show
results for $m_{\tilde q} = 300$~ GeV. It also shows the cosmological muon background
(black * line) energy distribution.}
\label{fig:evmulti}
\end{figure}

In Figure~\ref{fig:evmulti} we show the $\tau$ and $\mu$ rates in such telescopes, where we 
assume $\sqrt{F} = 10^{5}$~GeV, representing scenarios where all NLSPs decay.
The event rate per year for these extended volumes are shown in Table~\ref{tab:ev1y}.
The distribution for the $\tau$ and $\mu$ arrival energy at the detector  for these events are very similar to the ones shown in 
Figure~\ref{fig:endet}.

%We also show the number of events per year as a function of $\sqrt{F}$ for 21 and 65~Km$^3$
%neutrino telescopes in Figure.

As shown in Table~\ref{tab:ev1y} the $\tau$ and $\mu$ rates in extended neutrino telescopes
are significant, with a few events per year for $m_{\tilde q} = 300$~GeV and 1 event per year
in the larger volume for $m_{\tilde q} = 900$~GeV. These rates are at the same level as
NLSP direct detection~\cite{abc,abcd} and indirect dark matter detection~\cite{edsjo} 
in 1 Km$^3$ neutrino telescopes. Therefore one needs to determine possible backgrounds for these events. 

\subsection{\label{sec:bckg} Backgrounds in Multi-Km$^3$ Telescopes}

Under the scenarios we are considering, while NLSP direct detection has a clean and very
distinctive signature, with two NLSPs transversing the detector simultaneously, NLSP indirect 
detection has a considerable background. It comes both from atmospheric as well as 
from cosmological neutrinos, where the latter is also the source of NLSPs. In the
indirect case, each produced NLSP will decay at different times and therefore almost all
$\tau$s and $\mu$s will transverse the detector as a single event.

The number of events in detectors of different volumes is shown in Table~\ref{tab:ev1y}.
Figure~\ref{fig:evmulti} shows the energy distribution of cosmological
$\mu$s per year in both 21 and 65 Km$^3$ telescopes. Muons from atmospheric neutrinos have lower
energy compared to the cosmological neutrinos and can be significantly reduced. However
cosmological neutrinos
will yield an almost unreducible background. The only way to distinguish $\mu$s coming from
NLSP decays, would be through a time correlation with the NLSP produced $\tau$s. 

However there is also a small but significant number of $\tau$s for extended telescopes.
In the best case scenario, four taus per year can be seen in the 65~Km$^3$ volume telescope.
There are many techniques to distinguish $\tau$s from $\mu$s in neutrino telescopes, as described
in the next section.  Most of these signatures are favored by an increased telescope size. As
an example, in the case of the double-bang~\cite{dblbang} signature, two cascades can occur
inside the detector. For a 1 km$^3$ telescope there is an tight $\tau$ energy window, in
order to contain these cascades, while a larger detector has a much larger efficiency for
this signature. There is an ongoing effort to improve discrimination between
$\mu$s and $\tau$s in neutrino telescopes \cite{ictau}. Once this discrimination is done 
efficiently, the background for $\tau$s generated by NLSP decays will depend on neutrino oscillations.
If one assumes that cosmological neutrinos oscillates, always arriving in a $1:1:1$ ratio of
$\nu_e$,$\nu_\mu$ and $\nu_\tau$s, there will also exist an almost irreducible $\tau$ background
(see next section). In this case, $\mu$s and $\tau$s produced by NLSP decays have to be time correlated
in order to reduce the background. Otherwise, NLSP produced $\tau$s will be background free.

\section{\label{sec:tau} $\tau$ Signatures in Neutrino Telescopes}

In order to check our regeneration simulation, we generated $\tau$s from a distribution
of $10^6 \nu_\tau$s isotropically hitting the Earth. The regeneration process was then 
simulated as described in Section~\ref{sec:rate}. We analyzed the following $\tau$ 
signatures \cite{dblbang,tausg}: double bang, where the $\nu_\tau$ charge current interacts
in the detector producing a $\tau$ plus a hadronic shower and the produced $\tau$ 
subsequently decays producing a second shower; lollipop, where a $\tau$
enters the detector and decays inside it producing one shower; inverted lollipop,
where one shower is produced together with a $\tau$ which in its turn goes
through the detector, and finally the tautsi pop signature, where as in the double bang, two showers are
produced but too close to each other looking like a single shower.

\begin{figure}
%\vspace*{-1.5cm}
\includegraphics[width=\linewidth,height=7.5cm]{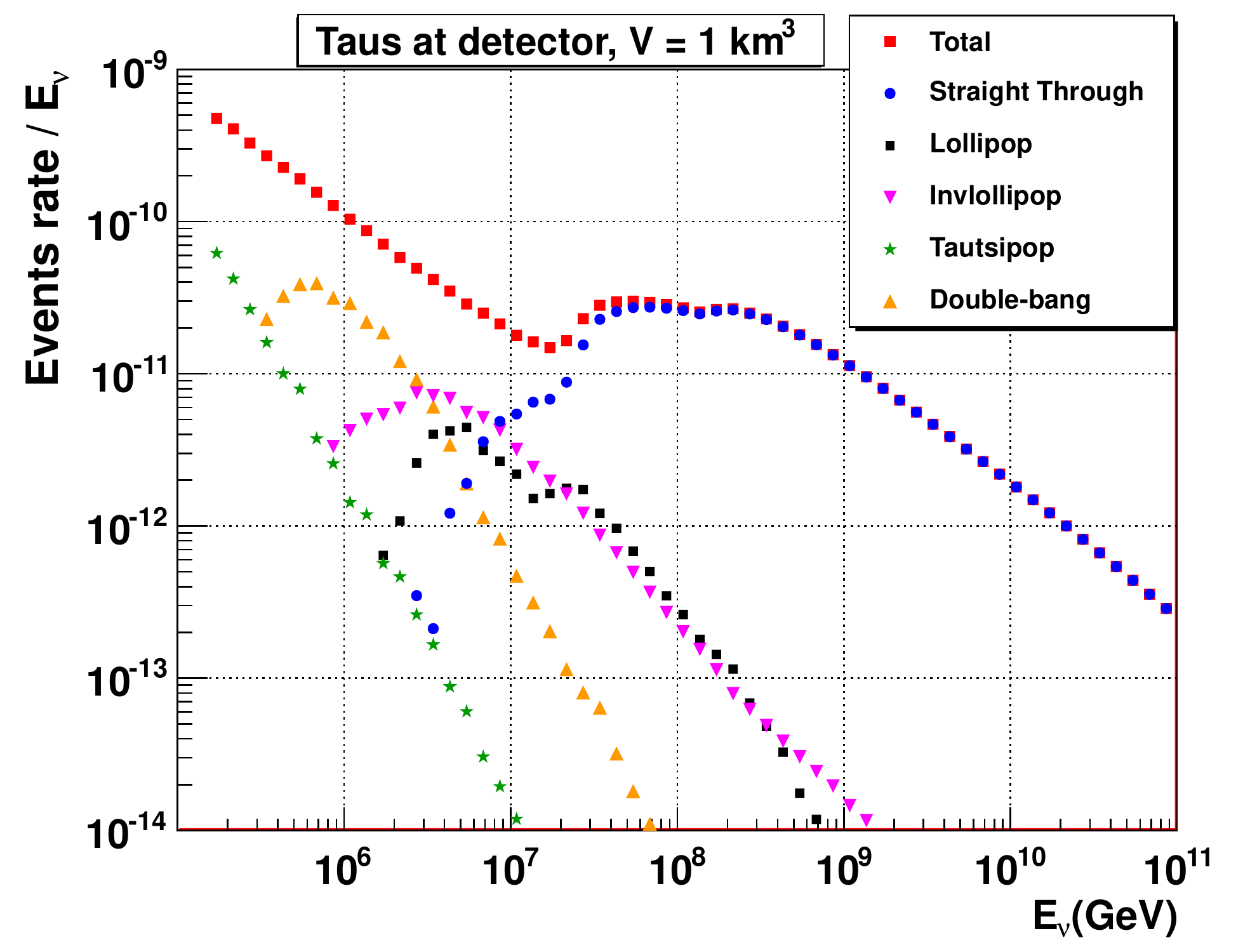}
\caption{Fraction of generated events which produce a specific $\tau$ signature as
labeled. In red is shown the total number of $\tau$s that reach the detector.
A 1 Km$^3$ volume is assumed.}
\label{fig:tausig}
\end{figure}

Figure~\ref{fig:tausig} shows the fraction of generated events
corresponding to each $\tau$ signature assuming a 1 km$^3$ volume, and the number of $\tau$s that 
go straight through the detector.  
Table \ref{tab:tausig} shows the fraction of events for each specific
signature, integrated in energy and the
number of events assuming that 1/3 of the WB upper limit corresponds
to $\nu_\tau$s. Both cases integrates the energy above $2 \times
10^6$~GeV, corresponding to the minimal energy with which most $\tau$s
arrive at the detector (see Figure~\ref{fig:endet}). These results are consistent
with \cite{termont}.

\begin{table}
\caption{Fraction of generated events with specific $\tau$ signature
  (see text), for 1, 21 and 65~Km$^3$ neutrino telescopes; and
integrated number of events assuming an initial $\nu_\tau$ flux equivalent to 1/3 of the WB 
limit. Only events with energies above $2 \times 10^6$ were accepted, since
most NLSP produced $\tau$s arrive the detector with energies above
this value.}
\vspace{10pt}
{\scriptsize
\hspace*{-.35cm}
\begin{tabular}{|c|c|c|c|c|c|c|}
\hline
\multicolumn{7}{|c|}{$\ $} \\
\multicolumn{7}{|c|}{V = 1 km$^{3}$} \\
\hline 
Limit   & Total & \emph{Lollipop} & \emph{Inv} &
\emph{Tautsipop} & \emph{Straight}  & \emph{Double} \\
           &          &     & \emph{Lollipop}  & & \emph{Through} &
             \emph{Bang} \\
\hline
$f_N (10^{-10})$   & 7.9  & 0.37  & 0.56  & 0.009  & 4.9  &  0.3 \\
\hline
WB       &   31    &    1.5  &  2.2 &   0.037  &   19.5  &  1.3 \\
\hline
\multicolumn{7}{|c|}{$\ $} \\
\multicolumn{7}{|c|}{V = 21 km$^{3}$} \\
\hline
$f_N (10^{-10})$   & 270  & 29  & 36  & 0.3  & 120  &  40 \\
\hline
WB       &   1052   &   114  & 143 &   1.3  &   457  &  158 \\
\hline
\multicolumn{7}{|c|}{$\ $} \\
\multicolumn{7}{|c|}{V = 65 km$^{3}$} \\
\hline
$f_N (10^{-10})$   & 860  & 83  & 130  & 1.2  & 300  &  200 \\
\hline
WB       &  3420 &   328   & 517   &  4.7   &   1180 &  790 \\
\hline
\end{tabular}
}
\label{tab:tausig}
\end{table}

Table~\ref{tab:tausig} shows that if 1/3 of the cosmological neutrino flux arrives at the
Earth as $\nu_\tau$, there will also be a significant cosmological background for the
NLSP generated $\tau$s. The only chance for extended neutrino telescopes to probe 
decaying NLSPs will therefore be a time correlation between both $\tau$s or a $\tau$ and
a $\mu$ produced from NLSP decay.

\section{\label{sec:conc} Conclusions}
We have shown that multi-Km$^3$ neutrino telescopes are potentially
sensitive to indirectly probe the scale of susy
breaking. However although a significant number of NLSP generated
$\tau$s will trigger these telescopes, this probe 
depends on the ability to distinguish these $\tau$s from a large
background. This background is composed of $\tau$s produced by
cosmological neutrinos, that will be significant if one assumes an
equal neutrino flavor ratio at the
Earth, due to oscillations. If for any reason oscillations do not
produce tau neutrinos, extended neutrino telescopes are able to probe
the susy breaking scale in the range $10^5 \lae \sqrt{F} \lae
10^6$~GeV. Large fluorescence telescopes can also indirectly probe \cite{aceuso}
the same $\sqrt{F}$ range but with the advantage of a background free
signature. Both these proposed indirect measurements complement direct susy breaking probes
that can be done by 1~Km$^3$ neutrino telescopes~\cite{abc,abcd}.

\acknowledgments
IA was partially funded by 
the Brazilian National Counsel for Scientific Research (CNPq), and J.C.S was  
funded by the State of S\~{a}o Paulo Research Foundation (FAPESP).


\begin{thebibliography}{99}
\bibitem{aceuso} Ivone~F.~M.~Albuquerque and Jairo Cavalcante de
  Souza,
%``Indirect Probes of Supersymmetry Breaking in the JEM-EUSO
%Observatory''
[arXiv:1209:43XX]
\bibitem{abc} I.~Albuquerque, G.~Burdman and Z.~Chacko,
  %``Neutrino telescopes as a direct probe of supersymmetry breaking,''
  Phys.\ Rev.\ Lett.\  {\bf 92}, 221802 (2004)
  [arXiv:hep-ph/0312197] 
\bibitem{abkk} I.~F.~M.~Albuquerque, G.~Burdman, C.~A.~Krenke and B.~Nosratpour,
  %``Direct Detection of Kaluza-Klein Particles in Neutrino Telescopes,''
  Phys.\ Rev.\  D {\bf 78}, 015010 (2008)
  [arXiv:0803.3479 [hep-ph]].
\bibitem{abcd} I.~Albuquerque, G.~Burdman and Z.~Chacko,
%``Direct Detection of Supersymmetric Particles in Neutrino Telescopes,''
  Phys.\ Rev.\  D {\bf 75}, 035006 (2007)
  [arXiv:hep-ph/0605120].
\bibitem{gmsb} G.~F.~Giudice and R.~Rattazzi,
  %``Theories with gauge mediated supersymmetry breaking,''
  Phys.\ Rept.\  {\bf 322}, 419 (1999)
  [hep-ph/9801271].
\bibitem{tese} J. C. de Souza, PhD. Thesis, 
  http://www.teses.usp.br/teses/disponiveis/,
  Universidade de São Paulo, in portuguese, São Paulo, 2012.
\bibitem{euso}   F.~Kajino {\it et al.}  [JEM-EUSO Collaboration],
  %``Overall view of the JEM-EUSO instruments,''
  AIP Conf.\ Proc.\  {\bf 1367}, 197 (2011);
  http:\/\/jemeuso.riken.jp/en/index.html
\bibitem{kmnet} Km3Net: http://www.km3net.org
\bibitem{icplus} F.~Halzen and D.~Hooper,
  %``IceCube-Plus: An Ultrahigh-energy neutrino telescope,''
  JCAP {\bf 0401}, 002 (2004)
  [astro-ph/0310152].
\bibitem{wb} E.~Waxman and J.~N.~Bahcall,
%``High energy neutrinos from astrophysical sources: An upper bound,''
Phys.\ Rev.\ D {\bf 59}, 023002 (1999); 
J.~N.~Bahcall and E.~Waxman,
%``High energy astrophysical neutrinos: The upper bound is robust,''
Phys.\ Rev.\ D {\bf 64}, 023002 (2001).
\bibitem{bbn}   M.~Kawasaki, K.~Kohri, T.~Moroi and A.~Yotsuyanagi,
 %``Big-Bang Nucleosynthesis and Gravitino,''
Phys.\ Rev.\ D {\bf 78}, 065011 (2008);
M.~Kawasaki, K.~Kohri and T.~Moroi,
 %``Big-Bang Nucleosynthesis with Long-Lived Charged Slepton,''
Phys.\ Lett.\ B {\bf 649}, 436 (2007)
\bibitem{gandhi} R.~Gandhi, C.~Quigg, M.~H.~Reno and I.~Sarcevic,
%``Ultrahigh-energy neutrino interactions,''
Astropart.\ Phys.\  {\bf 5}, 81 (1996); 
R.~Gandhi, C.~Quigg, M.~H.~Reno and I.~Sarcevic,
%``Neutrino interactions at ultrahigh energies,''
Phys.\ Rev.\ D {\bf 58}, 093009 (1998). 
\bibitem{earth} Adam Dziewonski, Earth Structure, Global, in The Encyclopedia of Solid Earth
Geophysics, edited by David E. James (Van Nostrand Reinhold, New York,
p. 331 (1989).
\bibitem{ina} M.~H.~Reno, I.~Sarcevic and S.~Su,
  %``Propagation of supersymmetric charged sleptons at high energies,''
  Astropart.\ Phys.\  {\bf 24}, 107 (2005)
  [arXiv:hep-ph/0503030].
\bibitem{crotty} 
  P.~R.~Crotty,
  %``High-energy neutrino fluxes from the supermassive dark matter,''
  FERMILAB-THESIS-2002-54 (2002).
\bibitem{dutta} 
  S.~I.~Dutta, M.~H.~Reno and I.~Sarcevic,
  %``Tau neutrinos underground: Signals of muon-neutrino ---> tau neutrino oscillations with extragalactic neutrinos,''
  Phys.\ Rev.\ D {\bf 62}, 123001 (2000)
  [hep-ph/0005310].


\bibitem{edsjo} L.~Bergstrom, J.~Edsjo and P.~Gondolo,
  %``Indirect detection of dark matter in km size neutrino telescopes,''
  Phys.\ Rev.\ D {\bf 58}, 103519 (1998)
  [hep-ph/9806293].
\bibitem{dblbang}
J.~G.~Learned and S.~Pakvasa,
  %``Detecting tau-neutrino oscillations at PeV energies,''
  Astropart.\ Phys.\  {\bf 3}, 267 (1995)
  [hep-ph/9405296, hep-ph/9408296].
\bibitem{ictau}
R.~Abbasi {\it et al.}  [IceCube Collaboration],
  %``A Search for UHE Tau Neutrinos with IceCube,''
  arXiv:1202.4564 [astro-ph.HE].
  %%CITATION = ARXIV:1202.4564;%%
\bibitem{tausg} D.~F.~Cowen [IceCube Collaboration],
  %``Tau neutrinos in IceCube,''
  J.\ Phys.\ Conf.\ Ser.\  {\bf 60}, 227 (2007).
  %%CITATION = 00462,60,227;%%
\bibitem{termont} E.~Bugaev, T.~Montaruli, Y.~Shlepin and I.~A.~Sokalski,
  %``Propagation of tau neutrinos and tau leptons through the earth and their detection in underwater / ice neutrino telescopes,''
  Astropart.\ Phys.\  {\bf 21}, 491 (2004)
  [hep-ph/0312295].
  %%CITATION = HEP-PH/0312295;%%
\end{thebibliography}
\end{document}